\documentclass[review]{elsarticle}
\usepackage{geometry}
\usepackage{url}
\usepackage[utf8]{inputenc}
\usepackage{amsmath}
\usepackage{amsfonts}
\usepackage{amssymb}
\usepackage{graphicx}

\begin{document}
	
\begin{frontmatter}
	
	\title{A theoretical approach to the nanoporous phase diagram of carbon}
	
	\author[a]{L.M. Mejía-Mendoza}
	\author[a]{M. Valdez-Gonzalez}
	\author[a,b]{Jesús Muñiz}
	\author[c]{U. Santiago}
	\author[a]{A.K. Cuentas-Gallegos}
	\author[a]{M. Robles}
	\address[a]{Instituto de Energías Renovables Universidad Nacional Autónoma de México, Privada Xochicalco S/N, Temixco, Morelos, 62580, México.}

	\address[b] {CONACYT-Universidad Nacional Autónoma de México, Privada Xochicalco S/N, Temixco, Morelos, 62580, México.}
	\address[c]{Departament of Physics and Astronomy, The University of Texas at San Antonio, One UTSA Circle, San Antonio, TX, 78249, USA.}

	
	\begin{abstract}

Nanoporous carbon has been extensively used in a wide range of applications ranging from water treatment to  electrochemical applications, such as in energy storage devices. An effort to relate structural to thermodynamical properties has not been explored from an atomistic approach. In this work we present numerical strategies to produce and study nanoporous carbon structures, using molecular dynamics simulations and a many-body potential. We designed a heating-quenching procedure in a thermodynamic region bounded by the critical, and triple point densities of carbon to study an ensemble of 1750 atomic arrangements produced at different densities, quench rates, and using graphite and diamond unit cells as precursor structures. All these samples were numerically characterized through the calculation of the free volumes, surface areas, radial distribution functions, and structure factors.
We found particularly useful the potential energy dependence with $sp^3$ hybridization content, to determine structural phases through clustering methods. Three phases were related to graphite-like, sponge-like, and unstable states. We showed that our results are compatible with available experiments and different theoretical schemes, concluding that the use of Tersoff potential is a reliable choice to produce nanoporous structures with low computational cost.
 
	\end{abstract}

	
\end{frontmatter}

\section{Introduction}
\label{Int}

Triple point of carbon, where graphite, liquid and vapor coexist, is known to occur at temperatures located between $4800\ K$ and $4900\ K$ and pressures around $108\ bar$ \cite{Savvatimskiy2015}. Therefore, in Earth's surface, where temperatures and pressures are around $300\ K$ and $1\ bar$ respectively, many disordered  atomic structures of carbon may exist in glassy-like metastable states. The nanoporous carbons are part of these states, and can be understood as amorphous solid structures containing hollow spaces with characteristic lengths below than $100\ nm$, high enough to allow diffusion of other atoms or molecules through them.

Amorphous carbons are experimentally defined in terms of the sp$^2$ and sp$^3$ ratio, where the sp$^3$ hybridization may be as high as 12 \% \cite{Robertson1986}. The properties of amorphous carbons are highly dependent on their nanoporosity due to their ability to adsorb atoms, combined with their electronic properties \cite{Kun2012,Legrain2015,Varanasi2015,Burt2014}, which are useful for many applications. Such as in air and water filters, and electrodes for supercapacitors and batteries \cite{Candelaria2012}.

The improvement of these energy storage devices, faces the challenges of understanding electrochemical processes like ion intercalation \cite{Ramireddy2014,Fong90} in lithium batteries  or electric double layer formation 
\cite{Burt2014,Chmiola2008} in supercapacitors. In these problems the use of numerical simulations play an essential role, due to the experimental difficulties that arise measuring the atomic environment in the nanopores.

The first challenge on the study of theoretical methods to predict nanoporous carbons, resides in how to build up a set of atomic arrangements, with structural properties that are close to real geometries. For this problem may be crucial the choice of the interaction potential, the simulation technique, and the possibility to include quantum effects.  

The Reverse Monte Carlo (RMC) technique \cite{Pikunic2003,Rosato1998}, has been used to obtain amorphous structures of carbon using the experimental radial distribution functions (RDF). The numerical procedure involve placing atoms at random positions in a compatible manner according to the RDF, then relax through a Tight Binding Molecular Dynamics (TBMD) simulation.  With RMC methods, we  can build structures of several thousand of atoms but, always restricted to use the experimental RDF or structure factor as input.

The \textit{ab initio} Molecular Dynamics (AIMD) \cite{Romero2010,Santiago-Cortes2012} simulation techniques are promising for the generation and study of amorphous samples. However they are computationally expensive since only systems of a few hundred of atoms can be generated in a reasonable calculation time. The search of an amorphous phase at this scale, is a very demanding task.

The use of Classical Molecular Dynamics (CMD) with the aid of many body potentials such as Brenner and Tersoff \cite{Brenner1990,Tersoff1994} is a reliable choice, and more recently,  with the introduction of Reactive Force Fields \cite{Chenoweth2008,Shi2008}, some quantum effects may be considered. With these alternatives, the use of CMD techniques are still being a source of new results concerning nanoporous carbon \cite{Ranganathan2017}. In previous works the numerical strategy to obtain the amorphous carbon has been usually based on melt-quenching methods. For instance, Peng et al. \cite{Peng2012} generated a set of amorphous carbon structures by variations of density and quenching rate to study the Hydrogen adsorption over them. The samples were built with densities ranging from 0.6 to 2.4~g/cm$^{3}$ using TB molecular dynamics. They found that the samples agree in some extent to the experimental radial distribution function (RDF) obtained by Gallego et al.\cite{Gallego2011}. Li et al. \cite{Li2013} found amorphous samples by quenching from the melt technique in the common allotropic carbon densities, i. e., densities ranging from 2.0 to 3.2~g/cm$^3$ and using a variety of empirical potentials such as REBO, ReaxFF and Tersoff \cite{Marks2002,Chenoweth2008,Tersoff1994}. It was reported that Tersoff potential is a suitable choice for carbon structures with density around diamond, that is close to 3.5~g/cm$^3$. Very recently Raganthan et al. \cite{Ranganathan2017} also employed this melt-quenching method within a CMD simulations using Reactive potentials, finding that at low densities the structures are predominantly anisotropic. At densities higher than 1~g/cm$^3$ the differences are due to the hybridization.   	

The aim of this work is to generate computationally nanoporous structures of carbon, proposing a simple and systematic numerical strategy, that allows to obtain an ensemble of hundreds of structures that may probably exists in nature. Our purpose is to search clues to define a structural phase diagram of nanoporous carbon. We sustain that the ensemble may bring new insights into the geometrical disposition of these structures, and their probable physical origins. Such properties may be the foundation for the creation of descriptors to design materials for carbon electrodes useful for energy storage applications.

We based our numerical study on some strategies taken from the phenomenon of bubble nucleation in liquids \cite{Uline2007}. This has been studied through Lennard-Jones fluids, using CMD or Monte Carlo (MC) simulations in superheated conditions \cite{Abascal2013,Gonzalez2014,Gonzalez2015}. One strategy to form such bubbles is to place the system in an unstable condition at densities and temperatures between the critical and triple points. In real systems such states lead to a liquid-gas coexistence phase, and in numerical simulations tend to create void formations in the liquid, assumed as the nucleation of nano-bubbles. Considering this process, our main hypothesis considers that by using a CMD simulation of particles interacting with a potential suitable for carbon, performed at densities between the critical and triple point and with temperatures below the triple point, an spontaneous nucleation of bubbles takes place in a solid-gas phase, analogous to that obtained in liquid-gas coexistence. After quenching is performed, it leads to the formation of pores. The critical and triple point densities have been determined at around $0.64\ g/cm^3$\cite{Leider1973} and $1.37~g/cm^3$ \cite{Haaland1976}, respectively, and the triple point temperature at around $4800\ K$.

It is important to consider that Tersoff potential is a suitable choice for searching amorphous nanoporous phases at low densities due to its bond order characteristics and versatility to deal with sp$^2$ and sp$^3$ bonding states \cite{Tersoff1994}. 

There is experimental evidence in the literature of nanoporous carbon  at extremely low densities to compare with, for example Zani et al. \cite{Zani2013} produced carbon foam layers with densities in the range 0.001-1.0~g/cm$^3$. It was reported via a Raman study interpretation that the samples they produced are nearly pure sp$^2$ network which contains topologically disordered graphite-like regions, as the resolution settled by the conditions of  the Pulsed Laser Deposition technique used in this work. Our objective is also to evaluate the predictive characteristics of Tersoff potential in the low density regime, by comparing structural properties of the simulated amorphous samples such as surface area, volume fraction, RDF with experimental available data. 

First we will show the numerical techniques to perform simulations routed to obtain the ensemble of nonporous structures; followed by the discussion of the results; then the predicted capacity of the obtained structures in comparison with some experimental results will be provided. Finally we will present the conclusions and contribution of our work to the state of the art within this field.

\section{Simulation strategy}
\label{Sim}

To obtain a statistically representative ensemble of nanoporous carbons, in a systematic way, it requires to perform hundreds of numerical simulations. In order to perform it on areasonable time, it is very important to achieve a balance between the details of the simulation model and the physical meaning of the results. In a recent work, Raganthan et al. \cite{Ranganathan2017}simulated a set of structures by using Molecular Dynamics with reactive potential at the ReaxFF \cite{VanDuin2001} level, demonstrating the capability of modeling amorphous carbons with melt-quench methods in a wide range of densities, from $0.5\ g/cm^3$ to $3.2\ g/cm^3$. It was concluded that the simulation box and quench rate (QR) play an important role in the structural features, and some of the obtained results are compatible with experiments. 

To design a full numerical strategy to go further, by evaluating hundred of cases obtaining a large variety of possible different densities and quench rates, three choices have to be taken into account: a reliable interaction potential efficient in CMD simulations, a set of initial conditions defined as precursor structures, and a thermal melt-quench process able to obtain the most variety of possible structures. The first choice was to use Tersoff potential instead of rective force fields, the reasons are: to perform the simulations faster, increasing the number of sampled structures, and to answer an unexamined question: how this potential reproduce the real nanoporous structures.

\subsection{Heating-quenching process.}

Looking for the bubble formation as mentioned in erlier section, we propose to examine densities between the critical an triple points. Also, the critical point density and temperature are found at $\rho_c=0.64\ g/cm^3$ and at $T_c=6800\ K$ \cite{Leider1973}, respectively, while the triple point density and temperature are found at $\rho_t=1.37\ g/cm^3$ and $T_t=4300\ K$ \cite{Haaland1976}, respectively. If we perform simulations from $T_a=300\ K$ to temperatures below  $T_t$, we guarantee the system does not melt and therefore we expect all samples to form bubbles, that after quenching lead to porous structures. As in glass production, and by the precursor initial configuration, as it is subsequently presented.  

The carbon precursor structures we examined were supercells with $6500$ atoms between $\rho_c$ and $\rho_t$, formed from: a) a graphite-like (GL), hexagonal based supercell that was generated by the 20$\times$20$\times$4 replication of an hexagonal unitary cell with the P6$_{3}$/mmm space group, and b) a diamond-like (DL), tetragonal supercell generated by 9$\times$9$\times$10 replication of a zinc-blende Fd$\bar{3}$m unitary cell. We set the density partition to be simulated in the range from $0.71\ g/cm^3$ to $1.37\ g/cm^3$ in steps of $0.02\ g/cm^3$ according to the following relation: 
\begin{center}
$\rho \in \Big\{0.71\textnormal{g/cm}^3+(0.02n)\textnormal{g/cm}^3 \Big\}^{33}_{n=0}$.
\end{center}

\begin{figure}[t] 
		\begin{center}
			\includegraphics[height=5cm]{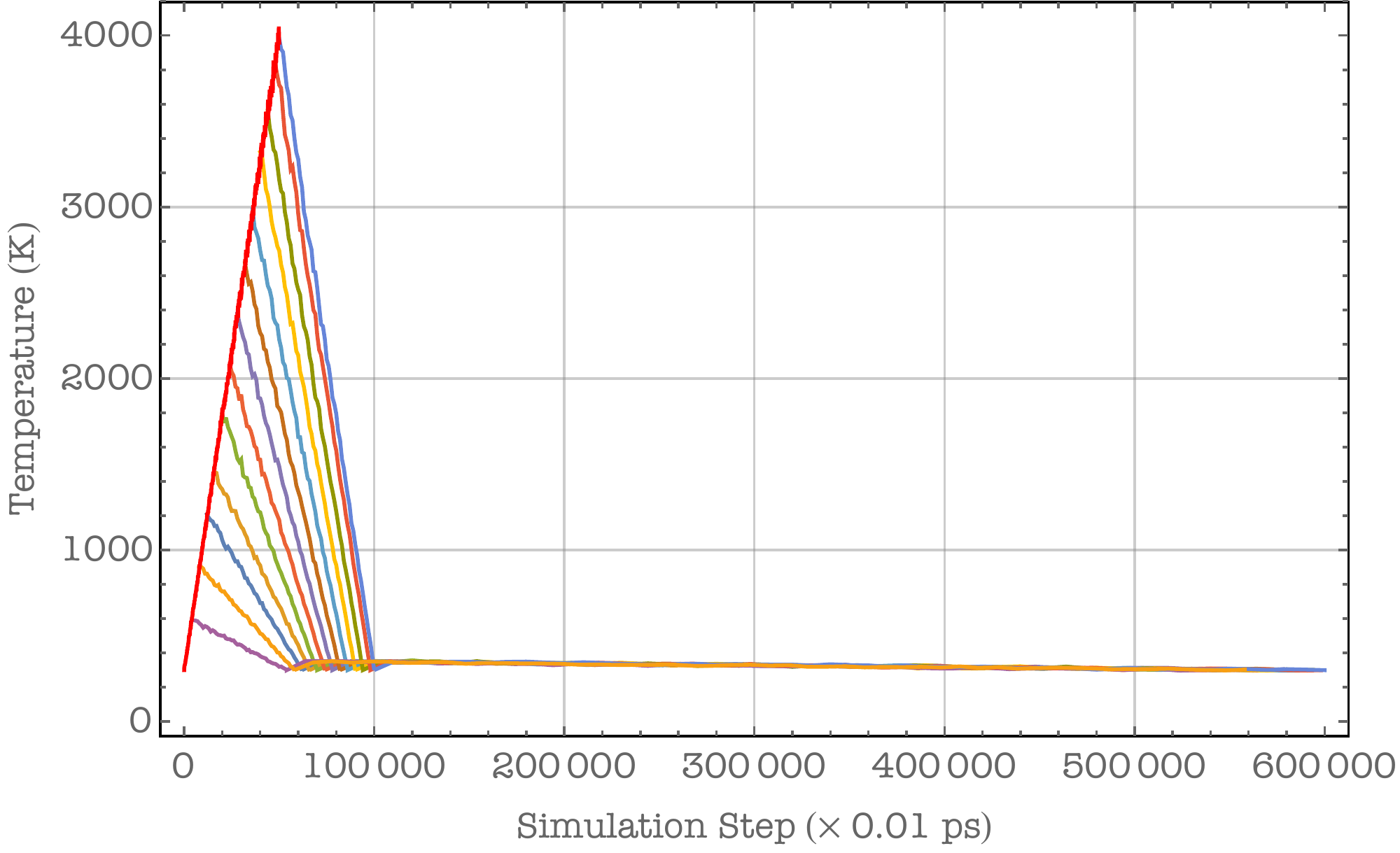}
			\includegraphics[height=5cm]{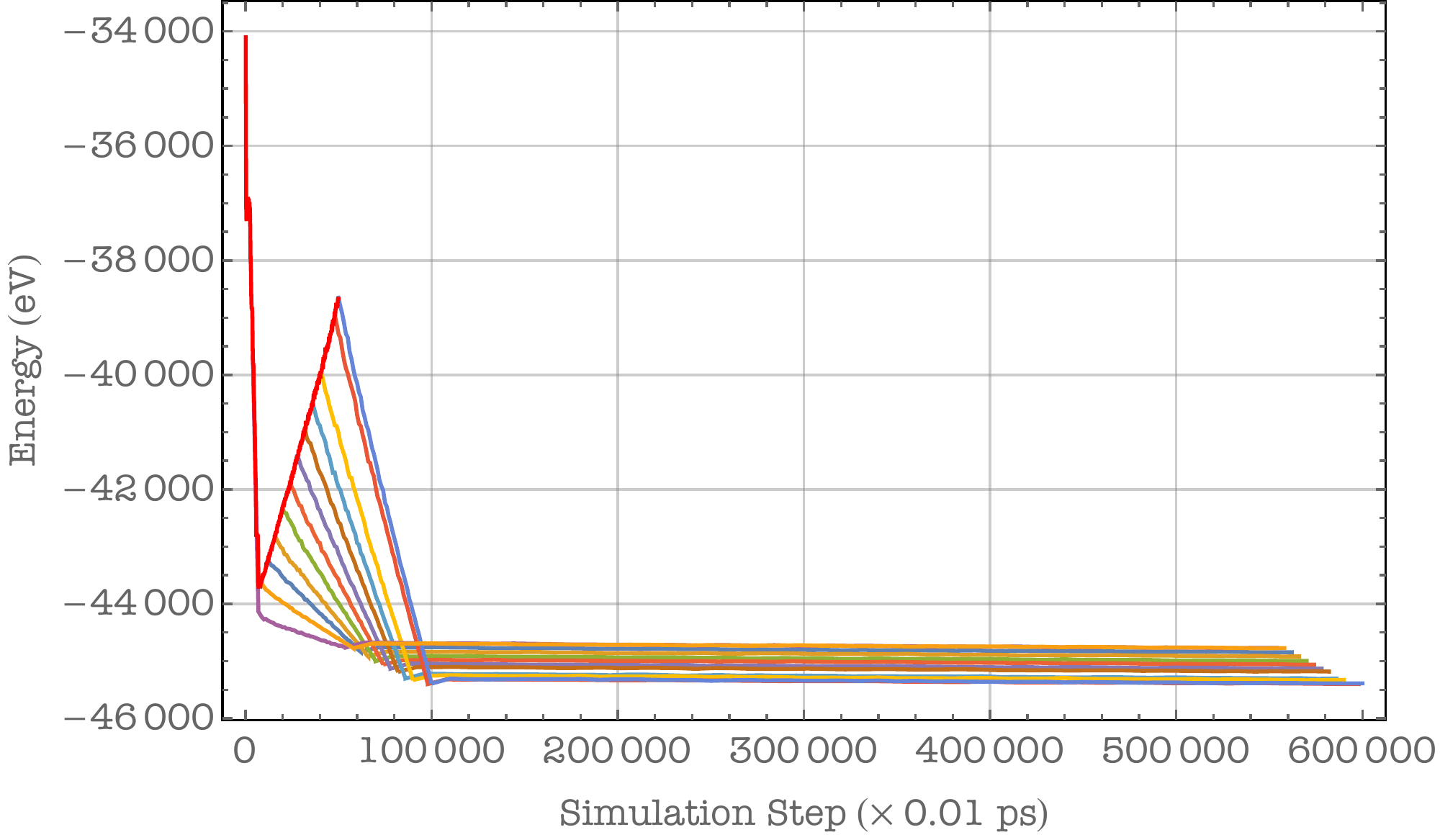}\\
			(a) \hspace{9cm} (b)
		\end{center}
		\caption{\small{a) \textit{Temperature vs Simulation step. Linear heating and quenching each 2000 simulation steps}. b) \textit{Total energy vs Simulation step, showing that the most relaxed structures are those with the fastest quenching rate}.}}
		\label{fig:1}
	\end{figure}

The simulations were performed using CMD in LAMMPS simulation package \cite{Plimpton1995} with its predefined Tersoff potential. The thermal ramps were set up as a linear heating from $300\ K$ to $4000\ K$ in $50000$ steps and using a time step of 1 fs, for all the examined densities. Every $2000$ simulation steps the actual positions an velocities are saved. Further, a quench is applied to cool down back to $300\ K$, producing 25 amorphous samples from each ramp, with quenching rates ranging from $2.96\times 10^{12}\ K/s$ to $74\times10^{12}\ K/s$. The volume was kept constant during the simulation by an isothermal-isobaric canonical ensemble and the Nosé-Hoover NVT thermostat was used  \cite{Nose1984, Hoover1985}. In order to stabilize each nanoporous structure, we applied a NVT process at $300\ K$ during $0.5\ ns$ after quenching, which certifies equilibrium in the novel amorphous carbon phases. The thermal procedure adopted and the total energy are sketched in Figure \ref{fig:1}.

\begin{figure}[t] 
		\begin{center}
			\includegraphics[width=9cm]{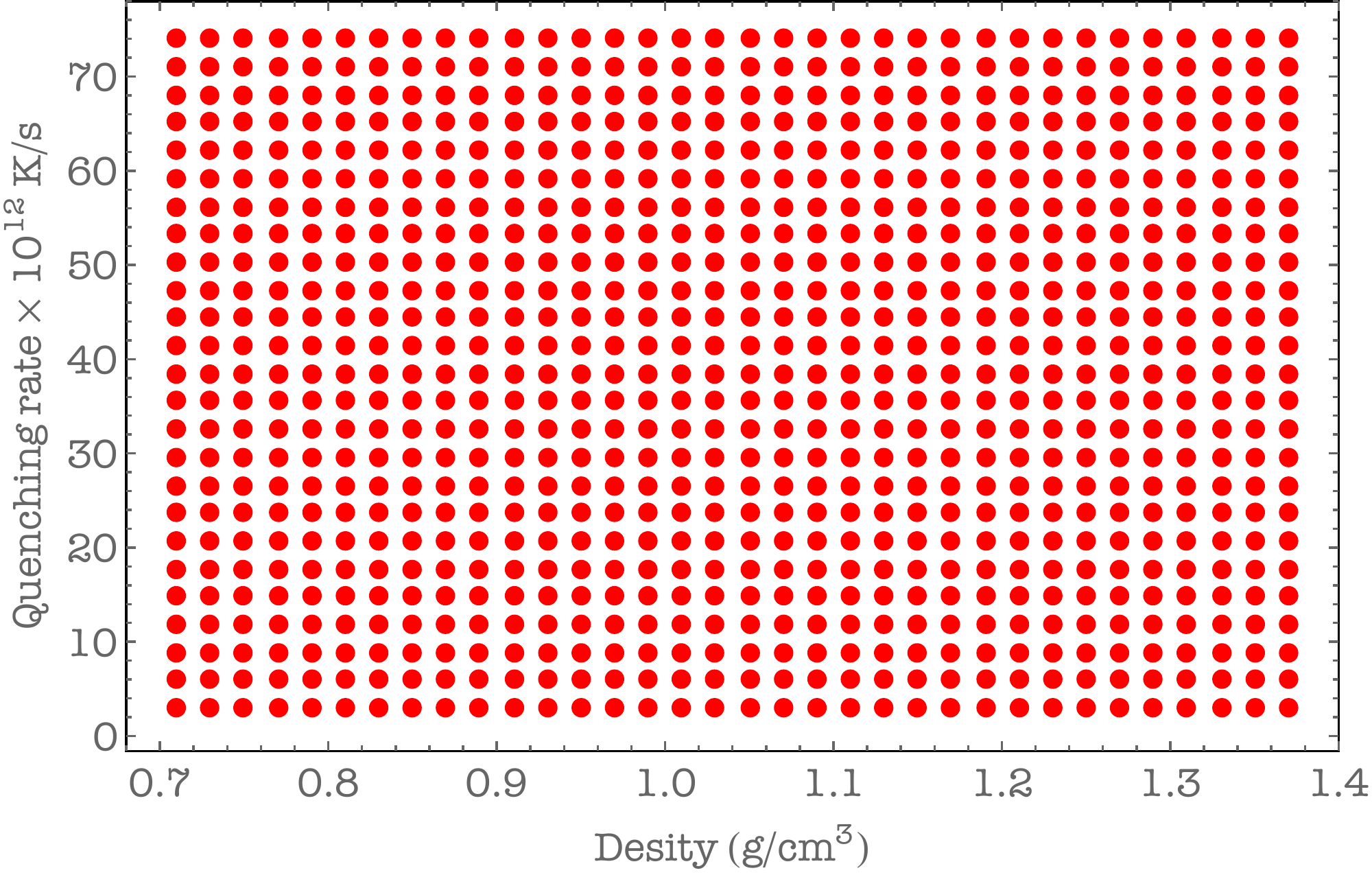}
		\end{center}
		\caption{\small\textit{Sampling space that represents the initial configurations for both structure sets, namely DL and GL}}
		\label{fig:2}
	\end{figure}

For the simulation conditions presented, the two controlled parameters; namely, density and quench rate, define the sample space depicted in Figure \ref{fig:2}. This sampling space represents the initial configurations settled in the same way for both precursor structure sets, DL and GL. The molecular dynamics heating procedure, described in the last paragraph, applied to this space, lead us to produce a set of $825$ different atomistic models of amorphous nanoporous carbons.

\subsection{Density Functional Theory (DFT).}

We set a parallel strategy at the quantum scale to benchmark our classical results with the previous reference of  Li et al. \cite{Li1990}. In that work the the authors compared among bond order potetentials as defined for Tersoff, reactive potential as defined for ReaxFF, Reactive Empirical Bond Order (REBO) potential and DFT methods of carbon structures generated via the melting-quench technique, with the following densities: 0.75, 1.03, 1.40, 1.75, 2.20, 2.60, 3.00, 3.20 and 3.5~g/cm$^3$, which in most cases are higher than that of $\rho_t$. 

Since the volume fraction of the structures remains constant as the size of the supercell decreases, we decided to reduce the size of the carbon samples from 6500 to 125 atoms, amorphizing with CMD with the same thermal procedure described in section \ref{Sim}, for the densities previously cited. The final structures were used as input for a geometry optimization calculation using the plane wave code \emph{pw.x} included in the Quantum ESPRESSO suite \cite{Giannozzi2009}. In order to find the energy optimized configuration, the supercells were calculated assuming no spin polarization. The GGA Perdew-Burke-Ernzehof \cite{Perdew1996} exchange and correlation density functional were used. Also, a norm-conserving, scalar relativistic pseudopotential is  generated via the Martins-Troullier methodology \cite{Martins1993}. The plane wave energy cutoff was settled at $55\ Ry$ which is  equivalent to $748\ eV$. The convergence threshold between each self-consistent step was $10^{-8}\ eV$ and the force threshold between each BFGS step was $10^{-4}\  \textnormal{Ha}/a_{0}$ equivalent to $5.14 \times 10^{-3} \textnormal{eV}/\textnormal{\r{A}}$. We used the $\Gamma$-point for reciprocal space integration due to the lack of symmetry in the real space.

\subsection{Numerical Characterization.}

The quest of a phase diagram of nanoporous structures requires to select some relevant physical quantities to characterize the obtained atomistic models. Moreover, it should be comparable to experiments and useful for applied sciences. 
We choose a basic set of structural parameters easy to calculate with nowadays standard algorithms, already contained in ISAACS software \cite{LeRoux2010}: volume fraction (VF), $sp^3$ content($sp^3$), Radial distribution function (RDF) and Structure factor (S(k)). All this parameters are experimentally measurable.

For us an important parameter, specially relevant for electrode applications, is the surface area (SA), which at atomic scale can be defined and calculated using the Connolly algorithm \cite{Connolly1983, Connolly1983a}. It allows to compute in three dimensions, the accessible surface to an atom in the carbon matrices by replacing the repulsive interactions with effective solid spheres, using the van der Waals radii. A standard probe atom is the Nitrogen molecule ($N_2$), which is also commonly used in the experimental measurements. The $N_2$ molecule is anisotropic, but a mean radius is around $1.85\ $\r{A} \cite{Hirschfelder1964,Batsanov2001}. For carbon a reasonable experimental value is $1.7$ \r{A} \cite{Batsanov2001}. We have calculated the SA of our samples, rolling up the probe sphere over the van der Waals surface of carbon Periodic boundary conditions were used, dividing the total mass of the carbon atoms inside.

Experimentally disordered structures are often characterized through electronic microscopy. Actually, the highest resolution images are obtained with the transmission electron microscopy technique (TEM). This visualizations may also be simulated from the theoretical structures using the multislice method of Cowley and Moodie \cite{Cowley1957}. We used the SimulaTEM software developed by Gomez et al.\cite{Gomez2010} to generate images of our samples in an efficient way. The introduction of this analysis can enhance the comparison with real structures.

\section{Results}
\label{Res}

\begin{figure}[ht] 
		\begin{center}
			\includegraphics[width=16cm]{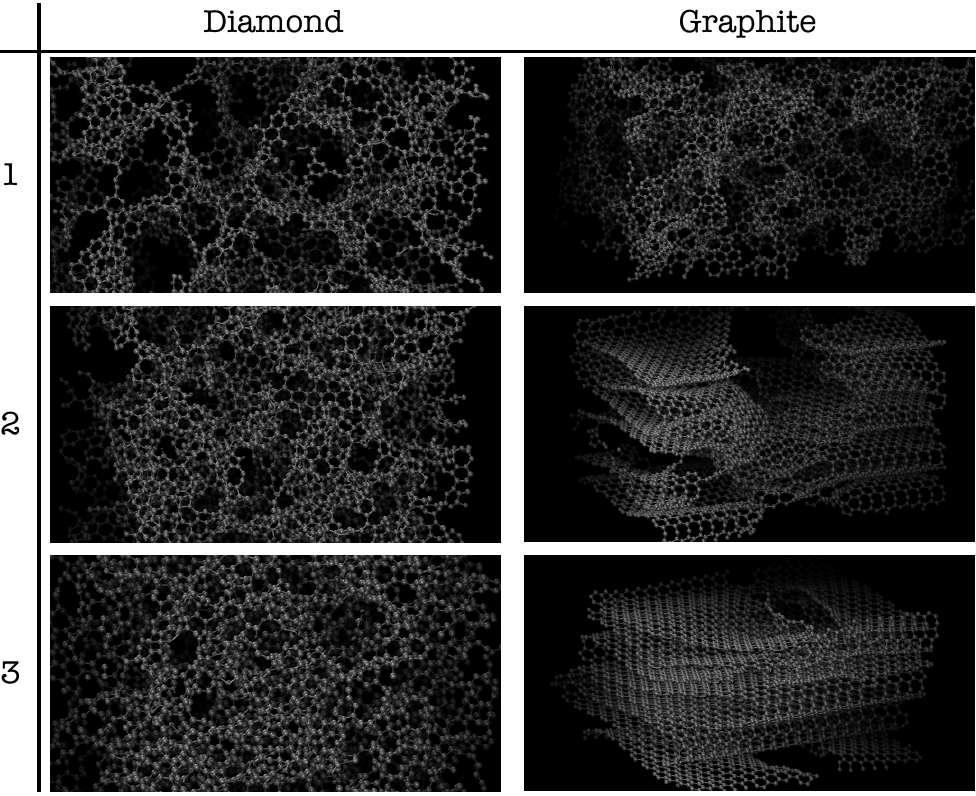}
		\end{center}
		\caption{\textit{Atomic positions of structures obtained with the heating-quenching procedure in three cases starting from graphite and diamond precursors: 1)  $\rho=0.71\ g/cm^3$ and  $QR=38.48\times10^{12}\ K/s$). 2) $\rho=1.03\ g/cm^3$  and $QR=35.52\times10^{12}\ K/s$. 3) $\rho=1.37\ g/cm^3$ and  $QR=41.44\times10^{12}\ K/s$.}}
		\label{fig:grid}
	\end{figure}

The applied heating-quenching procedure to graphite and diamond precursor states led to a set of $1750$ nanoporous structures. As an example, in Figure \ref{fig:grid} we selected the most common structures found at different densities namely, S1, S2 and S3, correspond to $0.71\ g/cm^3$  (with a quench rate of $38.48\times10^{12}\ K/s$), $1.03\ g/cm^3$ (at $35.52\times10^{12}\ K/s$) and $1.37\ g/cm^3$  (at$41.44\times10^{12}\ K/s$), respectively. It is important to highlight that at low densities both precursors show structures that resemble a sponge, which seems to be the favorite arrangement produced from diamond precursors;  meanwhile as we increase density, the graphite precursors tend to keep the layered order, but with the formation of voids.

\begin{figure}[t] 
		\begin{center}
			\includegraphics[width=9cm]{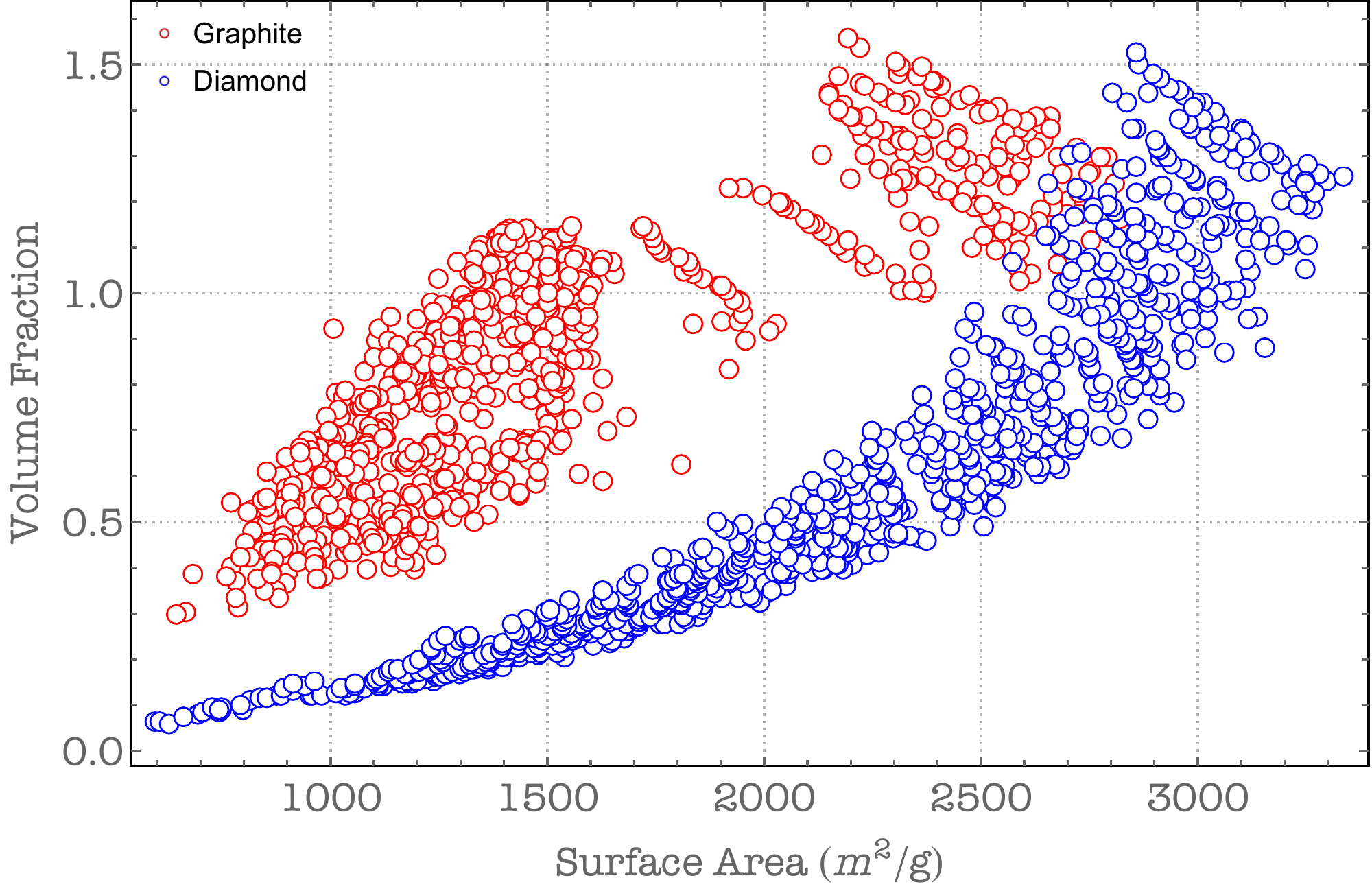}
		\end{center}
		\caption{\small\textit{Graphite (GL) and Diamond (DL) sets after the thermal procedure depicted in previous section.}}
		\label{fig:3}
	\end{figure}

The computation of the VF and SA allow to perform a characterization of all the ensemble of structures obtained after the thermal equilibration. The plot of SA vs VF, shown in Figure \ref{fig:3}, defines the regions where the produced structures lie. In red circles are shown the samples obtained from GL precursors and in blank circles are shown the DL group. Note that density is decreasing from left to right and in the lower density regime, which occurs when SA is over $1750\ m^2/g$, the points tend to line up as a function of quench rate, for both precursor sets. At higher densities, i.e. below $1750 m^2/g$, an apparent disordered clustering is raised, specially at the GL set.

\begin{figure}[ht] 
		\begin{center}
			\includegraphics[width=8.5cm]{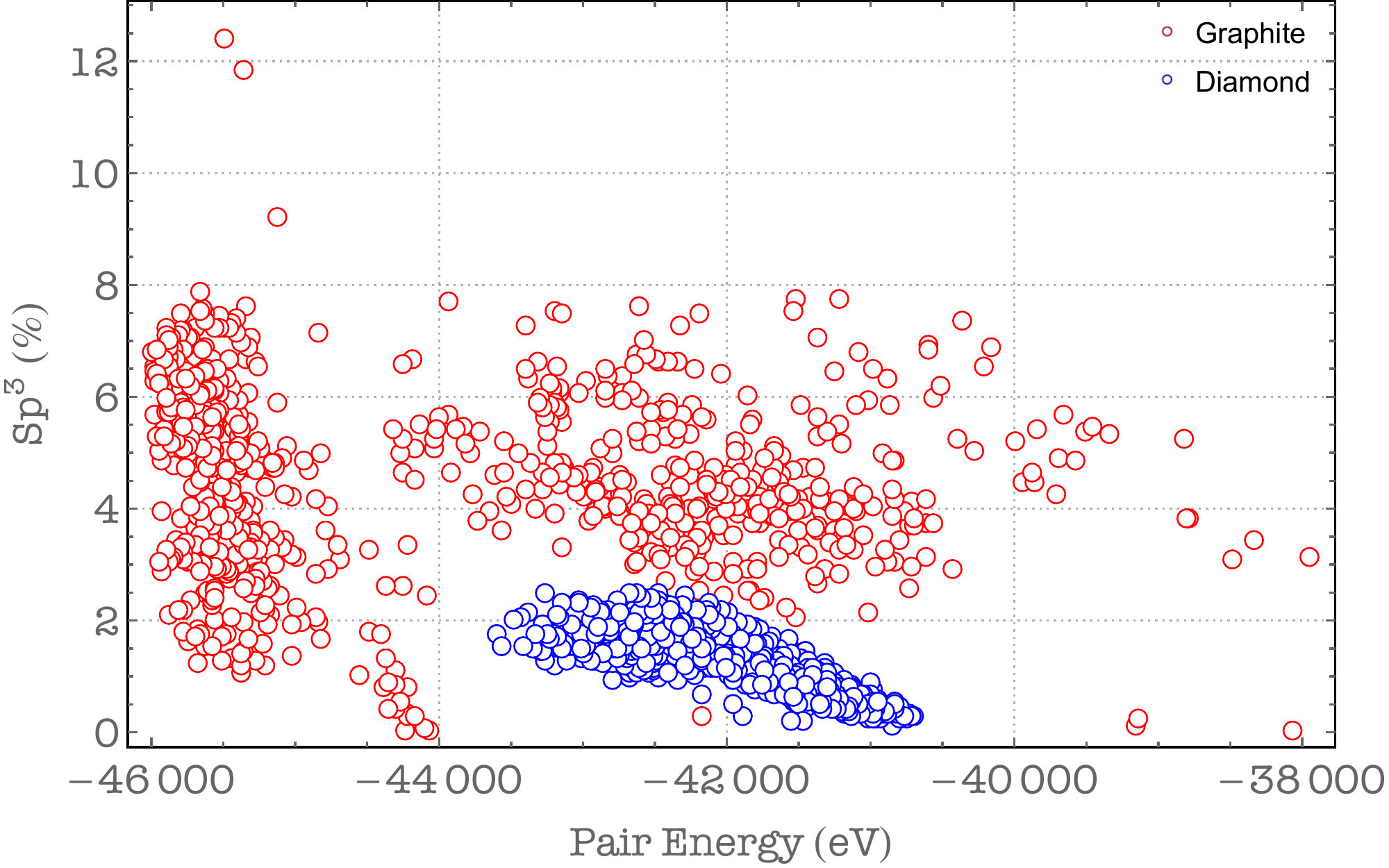}
			\includegraphics[width=8.5cm]{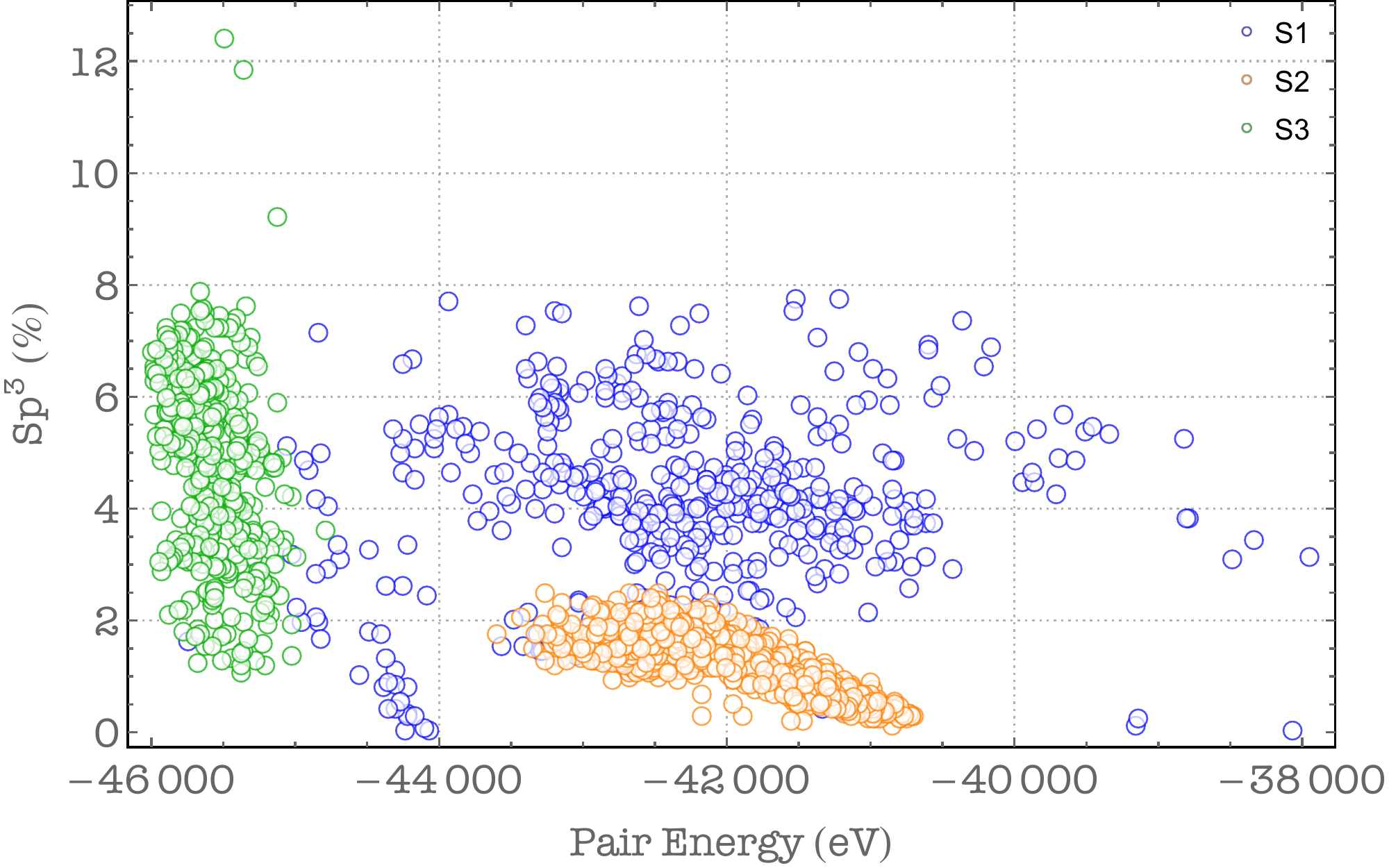}\\
			(a)\hspace{8cm}(b)
		\end{center}
		\caption{\small\textit{a) Values of $sp^3$ content versus pair energy. Note the \emph{natural clustering} of the DL structure set. b) The same distribution than that of a), classified by a Gaussian mixture clustering algorithm, where three main sets S1, S2 and S3 were obtained.}}
		\label{fig:4}
	\end{figure}

In order to go deeper in the structural analysis we decided to evaluate the $sp^3$ fraction and plotting it against potential energy (PE) to explore the energetic landscape, searching a way to identify and separate possible structural phases. The $sp^3$ fraction was measured with ISAACS method, which uses the following geometrical criterion: a carbon is hibridized with another carbon atom if the bond length is around $1.54$ \r{A}. In Figure \ref{fig:4}a we present the resulting PE vs $sp^3$ relation, for the complete set of structures divided in the two precursors GL (red) and DL (white). We must point out that the percentage of Sp3 content is in good agreement with the definition of an amorphous carbon \cite{Robertson1986}.

The resulting figure, shows that the GL structures separate in two main clusters while DL keep in a compact cloud. This suggest that data can be separated using simple computational clustering methods. A k-means algorithm  \cite{Hartigan1979} combined with a Gaussian mixture process, allow to automatically obtain the classification depicted in Figure \ref{fig:4}(b). It is worth to note that the DL main cluster, fits in a region between -43500 eV to -40500 eV and 0.1 \% to 2.8 \% . This set is labeled as S2. The two other sets belong almost entirely to the GL ensemble, and were labeled as S1 and S3. The energy extension of S1 is narrowly located between -46100 eV and -44600 eV and spreads out from 1 \% to 12.5 \%. Finally, the S2 set is sparcely located from -45000 eV to -37900 eV and 0.1 \% to 8 \%. The standard deviations from each set are 0.519, 1.685 and 1.710 for S2, S3 and S1 respectively.

\begin{figure}[htb] 
		\begin{center}
			\includegraphics[width=8.5cm]{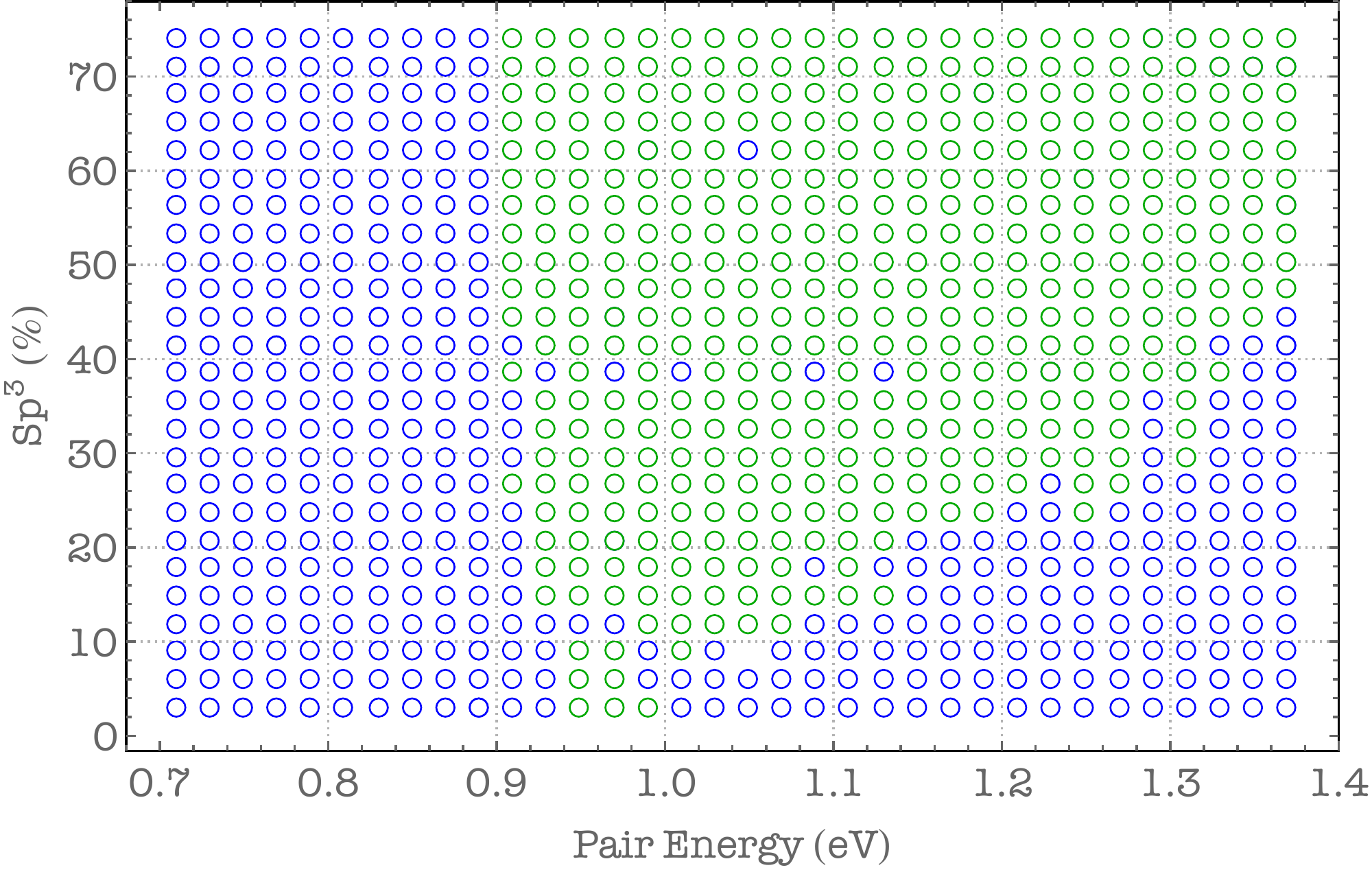}
			\includegraphics[width=8.5cm]{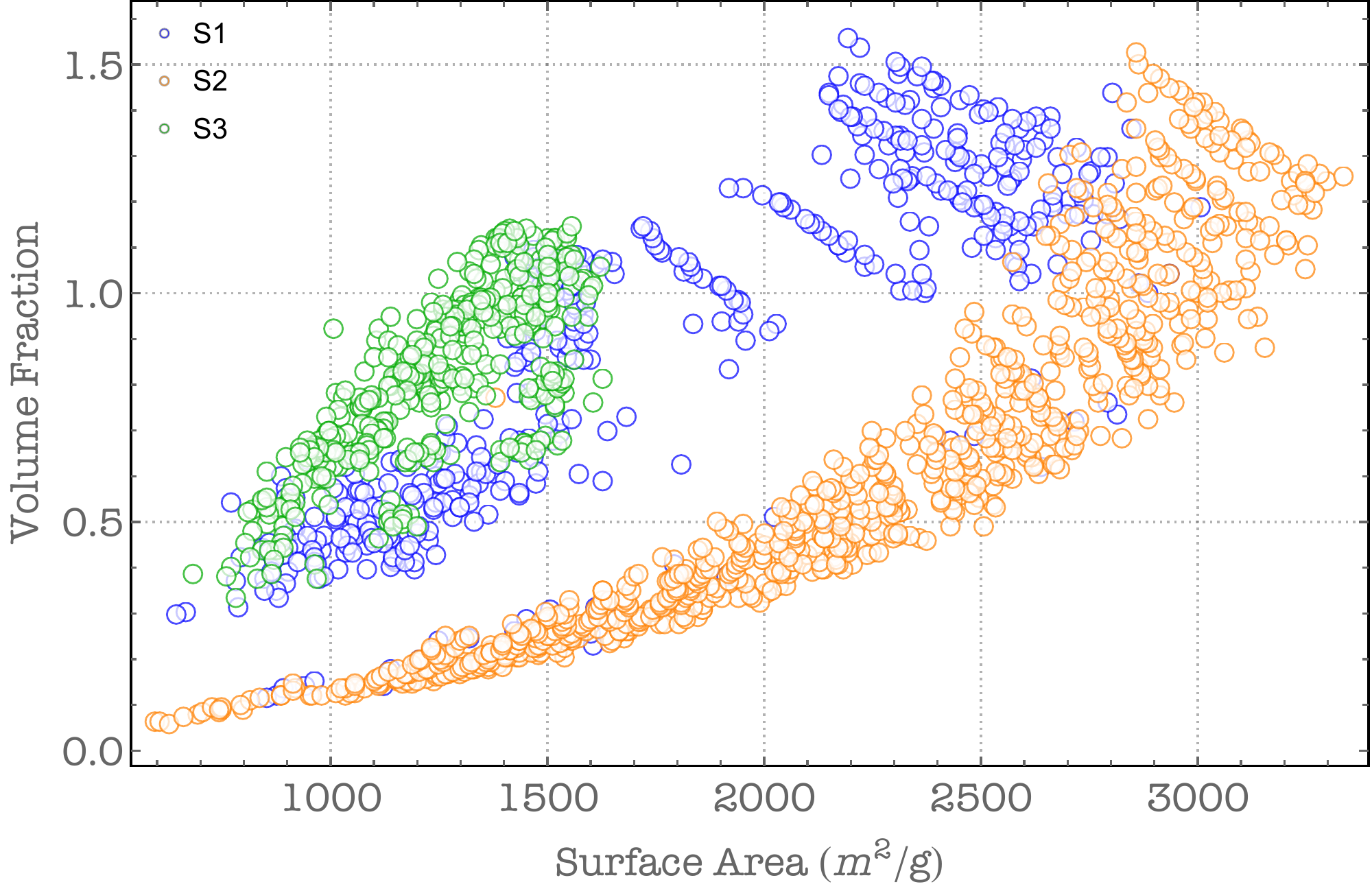}\\
			(a)\hspace{8cm}(b)
		\end{center}
		\caption{\small\textit{a) Projection of the clusters onto the original sampling space. Only S1 (blue) and S3 (green) sets are visible for the GL set. b) VF vs. SA diagram colored with automatic obtained clusters}}
		\label{fig:6}
	\end{figure}

We decided to visualize the classification S1, S2 and S3 in our sampling spaces: density vs. quenching rate and VF vs SA, in order to visualize possible phases in terms of the $sp^3$ fraction. These projections are depicted in Figure \ref{fig:6}(a) and (b). The two sets depicted, namely S1 and S3, in Figure \ref{fig:6}(a) shows that at low densities (up to 0.0436 \r{A}$^{-3}$) the distribution is made up of cluster S1, while above this density and for increasing quench rate, the predominant cluster is S3.

\begin{figure} [ht]
		\begin{center}
			\includegraphics[width=16cm]{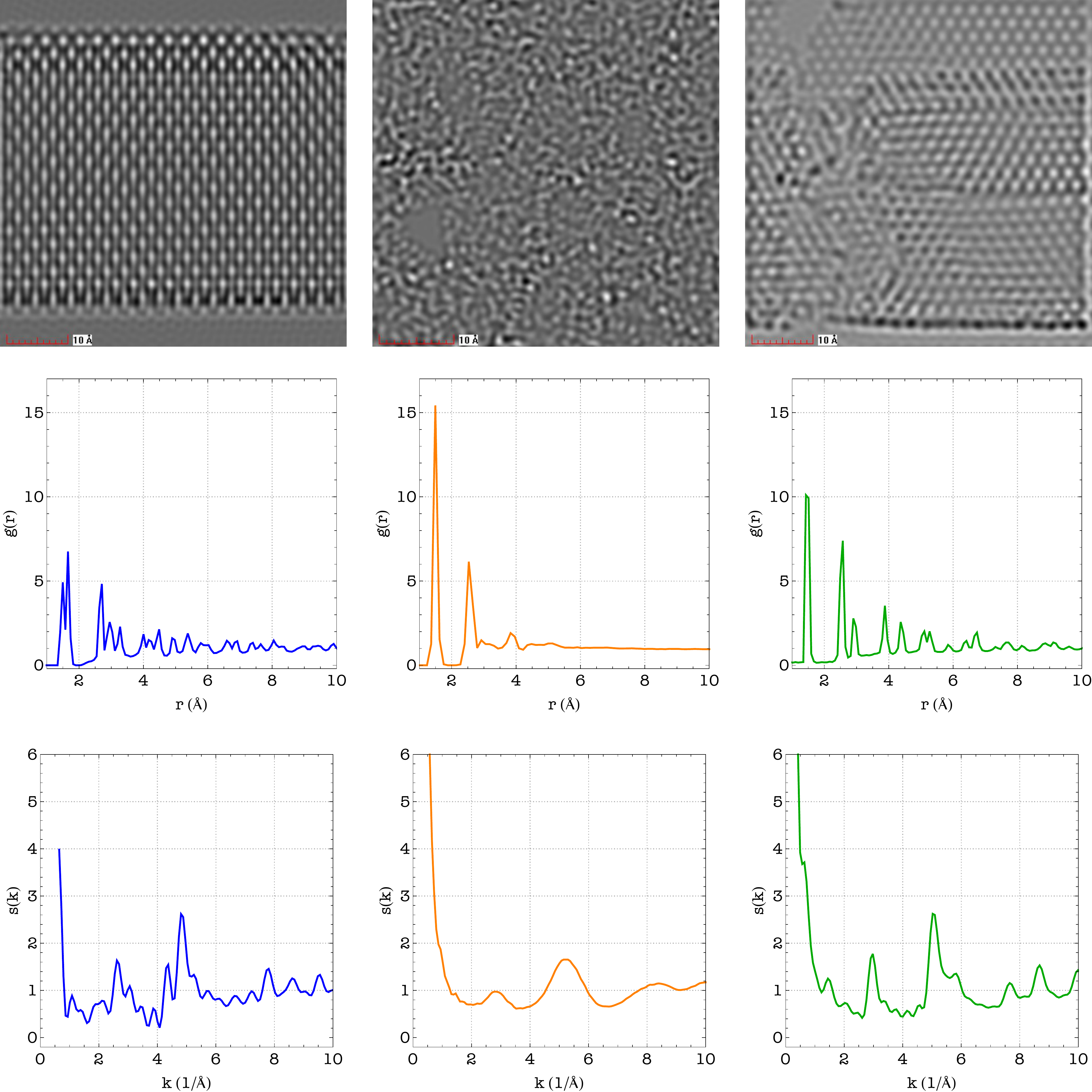}\\
			(a) \hspace{4.6cm} (b) \hspace{4.8cm} (c)
		\end{center}
		\caption{\small\textit{Comparison of the simulated TEM images, RDF, and s(k) between the three phases: (a) S1 high energy unstable states, (b) S2 sponge-like states (c) S3 graphite-like states. The corresponding coordination numbers for each phase are: 2.97, 2.79 and 3.19, respectively}}
		\label{fig:8}
	\end{figure}

We evaluated the RDFs, S(k) and their simulated TEM images generated by SimulaTEM software \cite{Gomez2010}, in order to find a relation between clusters and structural phases. As previously, the calculation of the S(k) was performed with the FFT algorithm implemented in ISAACS software \cite{LeRoux2010}. In Figure \ref{fig:8} we show samples from our three different sets already identified. In the cluster S1 we found different strucutres, which among other, like the example in the figure, are stressed graphite. In figure \ref{fig:8} (a), the RDF shows that the first maximum is composed of two peaks, one around 1.49 \r{A} and the other located at 1.65 \r{A}. The first position corresponds to the presence of dangling bonds and $sp^2$ carbon atoms \cite{Fitzer1995} where the coordination number its around 2, while the latter correspond to adjacent carbon atoms with no bonds, tracing out a highly stressed-high energy structure of carbon. The coordination number integrated to the first minimum of the RDF is 2.97. All this can be seen in the simulated TEM image, where the structures lying  in the interval are build up from distorted benzene-like rings. We also identified that structures which density lies in the interval from 0.71 to 0.87~g/cm$^3$ have structures similar to those in S2. The S1 cluster can still be formed by many unstable phases which present a small probability of existence.

The S2 samples, fig \ref{fig:8} (b), can be called a sponge-like phase, containing carbon atoms bonded at a distance of 1.49 \r{A} due to the existence of $sp^2$ and dangling bonds; the coordination number equals to 2.79 suggesting reinforcing the idea of the existence of dangling bonds.  The shape of the S2 structure factor is very smooth compared with the S(k) of the other two phases due to the atomic disorder which can be observed in the TEM image generated. 

In S3 set, we found graphite-like structures where the first coordination shell is located at 1.47 \r{A} with a coordination number equal to 3.19, suggesting that the number of dangling bonds decreases as compared with the other two phases. Here the graphite-like structures have rings of five, seven and eight members as can be seen at the SIMULATEM image of these phase. Since the energy of this structures is the lowest, we recognize this cluster as representing the most stable phase  

\begin{figure}[tb]
		\begin{center}
			\includegraphics[width=9cm]{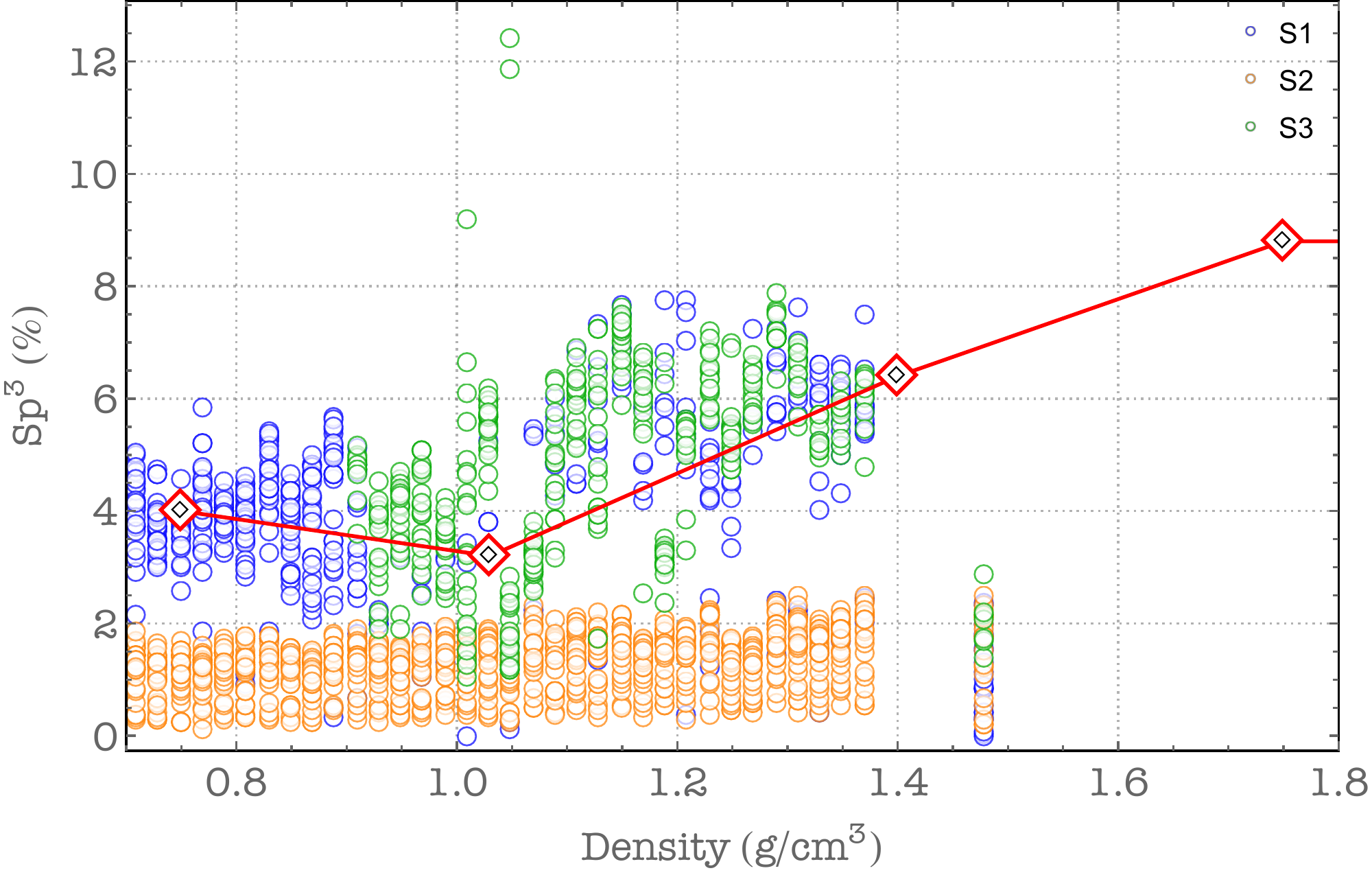}
		\end{center}
		\caption{\small\textit{Percentage of $sp^3$ vs. Density. In circles we show the Tersoff based structures belonging to Si, S2 and S3 sets, in red diamonds we show the results of DFT optimizations.}}
		\label{fig:10}
	\end{figure}

Finally, in order to have a DFT benchmark we generated porous structures from cubic samples of 125 atoms with the following densities: 0.75, 1.03, 1.40, 1.75, 2.20, 2.60, 3.00, 3.20 and 3.5~g/cm$^3$ using both, Tersoff potential and DFT relaxation. In Figure \ref{fig:10} we present the $sp^3$ percentage vs. density calculated for the densities 0.75, 1.03, 1.4 and 1.75~g/cm$^3$ and the whole sets S1, S2 and S3. We show that the quantity of $sp^3$ content vs. density lays in a region bounded by the values of the same calculation of the Tersoff's calculated S1, S2 and S3 sets. In next section, we want to describe and discuss the potential predictive properties of the diagrams described in this section.

\section{Discussion}
\label{Disc}
\subsection{Surface area, quench rate and Structure factors}

\begin{figure} [ht]
		\begin{center}
			\includegraphics[width=9cm]{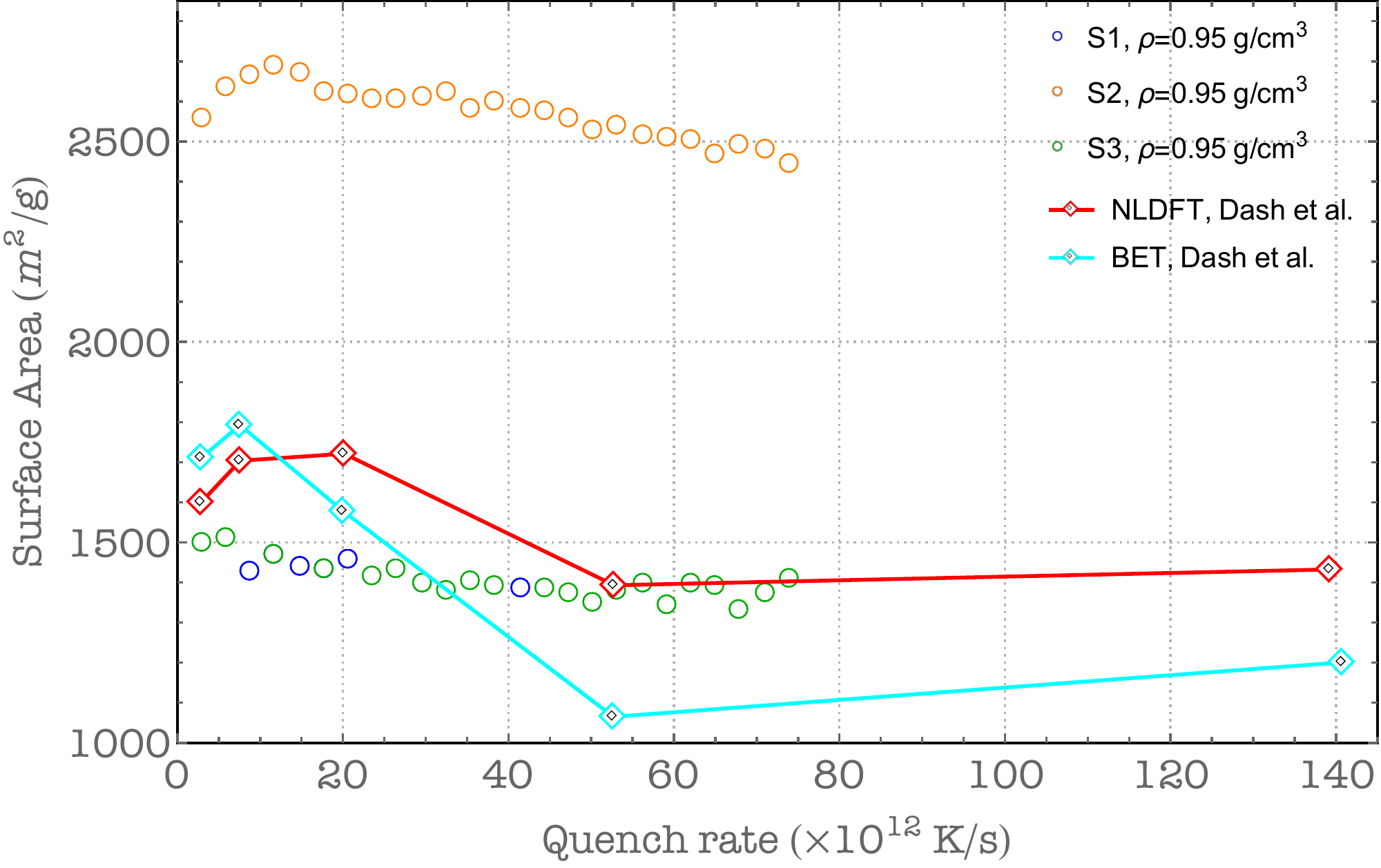}
		\end{center}
		\caption{\small\textit{Surface area vs quench rate for S1, S2 and S3 for $\rho=0.95~g/cm^3$ compared with those experimental values calculated by Dash et al. }}
		\label{fig:11}
	\end{figure}

 First of all, we can compare the results of Dash, et al. \cite{Dash2006} with our Tersoff results. In that work, the authors prepared a carbide-derived carbon, from TiC synthesized at temperatures, ranging from 400 to 1200~$^\textnormal{o}$C. Using the results of the analysis made by Palmer et al. \cite{Palmer2010},  which lays in the comparison of structural features among simulated structures and those of the carbide-derived-carbon reported by Dash et al., the synthesis temperature experimentally measured, can be related to a simulated quench rate. Therefore, we can compare our results with the measurement of the surface area done with NLDFT and BET techniques in nitrogen. We found that the best density that fits the experimental results is $0.95\ g/cm^3$, in coincidence with Palmer's simulation. estimation. In Figure \ref{fig:11} are shown the NLDFT and BET curves from Dash et al. work and the SA values of our samples at $\rho=0.95\ g/cm^3$, contained in S1, S2 and S3 clusters. In this figure we highlight that our measurements of SA vs. quench rate in the range from 45 to 74~$\times 10^{12}$~K/s belongs only to the S3 cluster or the low energy graphite-like nanoporous phase, with a reasonable agreement with the experimental results. In average, for the S2 set, the values of surface area are around 1.83 times greater than those from S1 and S3. In this case, it appears that the sponge-like structures are absent in this experimental setup.

\begin{figure} [ht]
		\begin{center}
			\includegraphics[height=6cm]{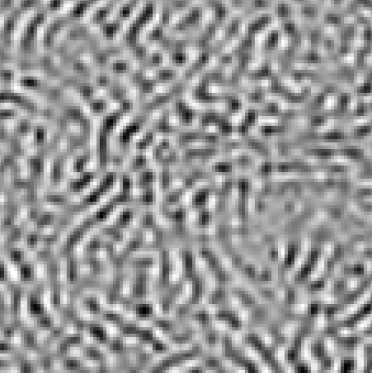}
		\includegraphics[height=6cm]{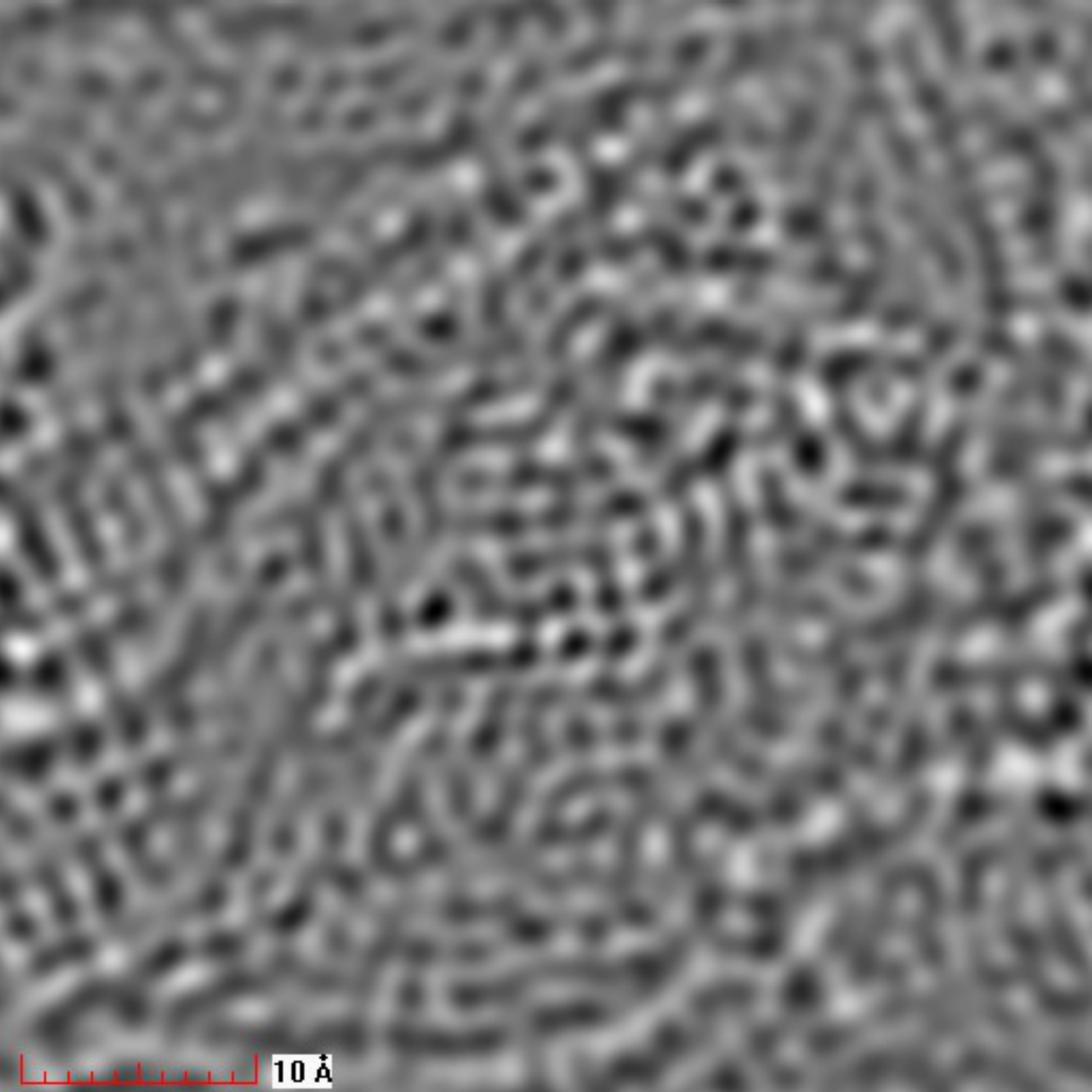}
			\includegraphics[height=6cm]{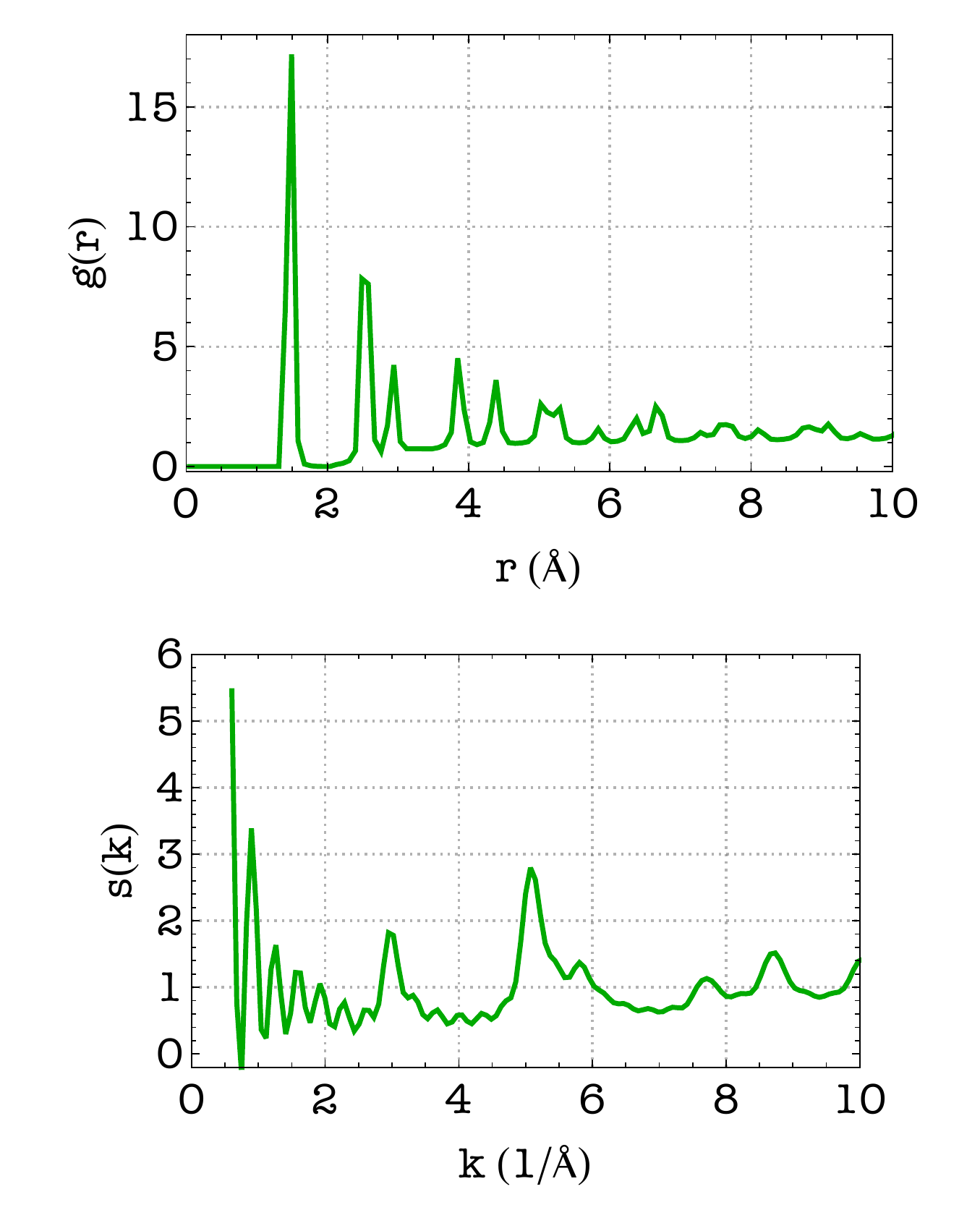}\\
			\hspace{1cm} (a) \hspace{5cm} (b) \hspace{5.5cm} (c)
		\end{center}
		\caption{\small\textit{(a) HRTEM image of nanoporous carbon synthesized at 800~$^\textnormal{o}$C from \cite{Dash2006}. (b) SIMULATEM image of the simulated graphite-like structure with $\rho=0.95$~g/cm$^3$ and $QR=50 \times 10^{12} K/s$. (c) RDF and S(k) calculated from the graphite-like structure, the both curves represents an amorphous nanoporous sample with some kind of graphite structure.}}
		\label{fig:12}
	\end{figure}

In figure \ref{fig:11}, , the best agreement between theoretical and experimental data corresponds to the BET measurement at a surface area of 1382~m$^2$/g. The corresponding theoretical structure has a QR equal to 53.28$\times 10^{12}$ K/s and belongs to the S3 set. For this case, we calculated the RDF, S(k) and TEM image, the results are in Figure \ref{fig:12}. The simulated TEM image of the Graphite like S3 (Figure \ref{fig:12}(b)), shows features of graphene-like sheets intercalated, similar to the HRTEM image, which corresponds to an intermediate temperature of 800 $^\textnormal{o}$C from Dash \textit{et. al} uch behavior is shown in Fig. \ref{fig:12} (a) shows. In Figure \ref{fig:12} (c) we have calculated the RDF for the sample depicted in Figure \ref{fig:12} (b) and shows that the first coordination shell is located at 1.49 \r{A} which corresponds to graphite like structures with the presence of dangling bonds \cite{Fitzer1995} and a coordination number of 2.93. The second coordination shell is located around 2.53 \r{A}, which corresponds to an intermediate position between the second peak in graphite, and the second peak in diamond. The third peak is located at 2.94 \r{A} and is related with the third peak in graphite at 2.93 \r{A}. Gathering the information contained in the simulated TEM image and the RDF analysis, we can suggest that the structure depicted here has a strong graphitic structure. Finally, from the S(k) it appears a noisy structure due to the lack of long range order and this does not resemble the graphite peak located around 2~\r{A}$^{-1}$, also our calculated peaks at 3 and 5.2~\r{A}$^{-1}$ are close to the maximum from the graphite S(k)(see supplementary material \cite{Supmat}), reinforcing that there is some kind of short range graphitic structure in this sample.

\subsection{Searching carbon structures.}

\begin{figure} [ht]
		\begin{center}
			\includegraphics[width=8.5cm]{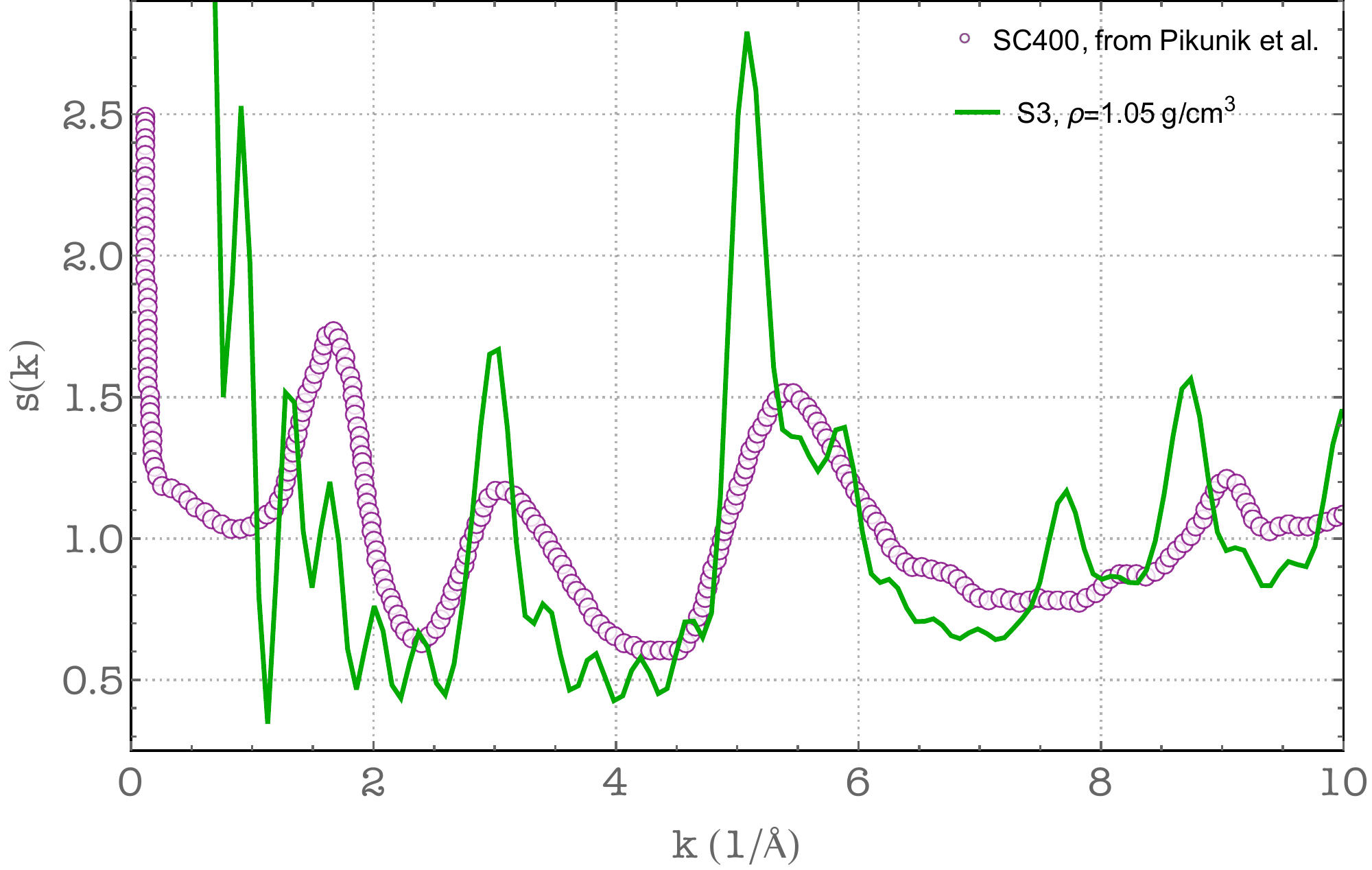}
			\includegraphics[width=8.5cm]{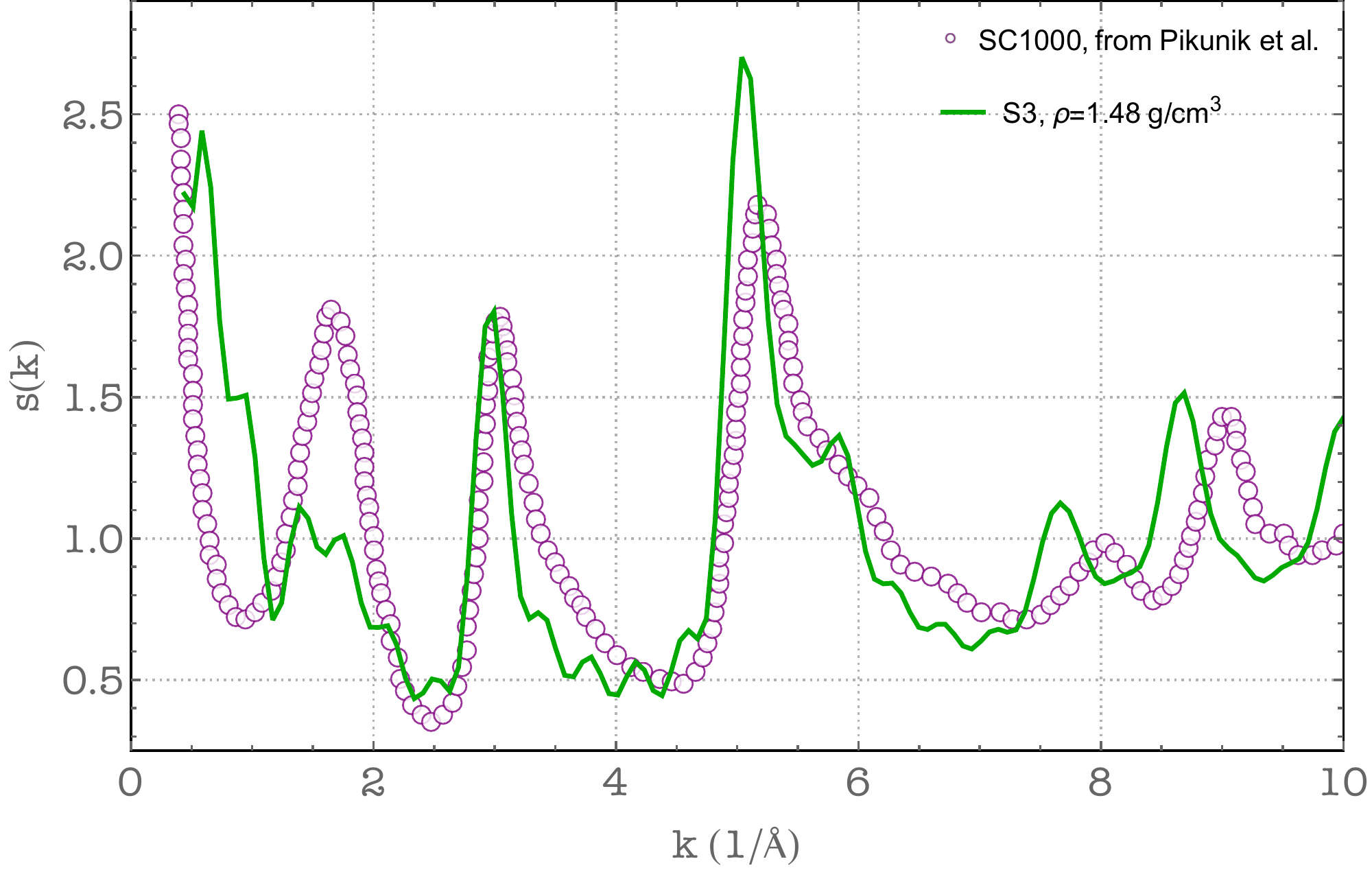}\\
			(a)\hspace{8cm}(b)
		\end{center}
		\caption{\small\textit{(a) Comparison between the structure factor experimentally derived from pyrolized SC400 sample (purple) and our S3 simulated sample of $\rho=1.05~g/cm^3$  (green). (b) Comparison between the experimental structure factor of SC1000 (purple) and our S3 sample of 1.48~g/cm$^3$ (green).}}
		\label{fig:13}
	\end{figure}

Another possible application for the collection of simulated structures, is to correlate the theoretical S(k) with those obtained experimentally. For example, in the work made by Pikunic \textit{et. al}  \cite{Pikunic2003}, reported the generation of nanoporous carbon using pyrolysis techniques applied to saccharose-based carbons in order to obtain two samples synthesized at 400 and 1000 $^\textnormal{o}$C. In this work, the authors reported a Reversed Constrained Monte Carlo method for the simulational construction of the Structure factor, which where done for the two synthesized samples. An alternative method is to search compatible structures in our simulated ensemble. We have chosen the two saccharose samples, reported by the authors as CS400 and CS1000 (pyrolyzed at 400 $^\textnormal{o}$C and 1000 $^\textnormal{o}$C, respectively). A systematic comparison with simulated S(k) led us to Figure \ref{fig:13}. The best choice of S(k) belongs to the S3 samples of QR equals to 74 $\times 10^{12}$ K/s and densities equals to 1.05 and 1.48~g/cm$^3$, respectively.

 Figure \ref{fig:13} (a) shows the experimental and simulated structure factors, for the CS400 sample. In the range from 1 to 2~\r{A}$^{-1}$ we observe, at the experimental curve, a peak around $k=1.7~$\r{A}$^{-1}$, due to long range structure; our calculated S(k) shows noisy behavior at this k, due to the introduction of a cut off in the RDF around 29 \r{A}, which defines a distance inside the supercell of the S3 atomic structure. The experimental peaks located at $k=3.1$ ~\r{A}$^{-1}$ and $k=5.5$  ~\r{A}$^{-1}$ are characteristic peaks of graphite-like structure, our simulated peaks around 3.0 and 5.1 ~\r{A}$^{-1}$ corresponds to the same structure; the simulated shoulder located at 5.7 ~\r{A}$^{-1}$ corresponds to diamond-like structure. At long range, our simulated peaks are shifted to the left due to the fluctuations at small distances, see \cite{Supmat}.

The comparison of the S(k) from CS1000 case and our S3 graphitic like structure are depicted in Figure \ref{fig:13} (b), we see that the first simulated peak is also lower than that of experiment, due to the cutoff distance in the RDF. The simulated graphite-like peaks are in agreement with our simulation at 3.0 and 5.1 ~\r{A}$^{-1}$, also the shoulder at 5.7 ~\r{A}$^{-1}$ coincides. The left-shift of the two last simulated peaks respect the experimental ones are due to fluctuations in the first distances in the RDF. Nevertheless the comparison is qualitatively and quantitative good.

\subsection{The sp$^3$ content.}

\begin{figure} [ht]
		\begin{center}
			\includegraphics[width=8.5cm]{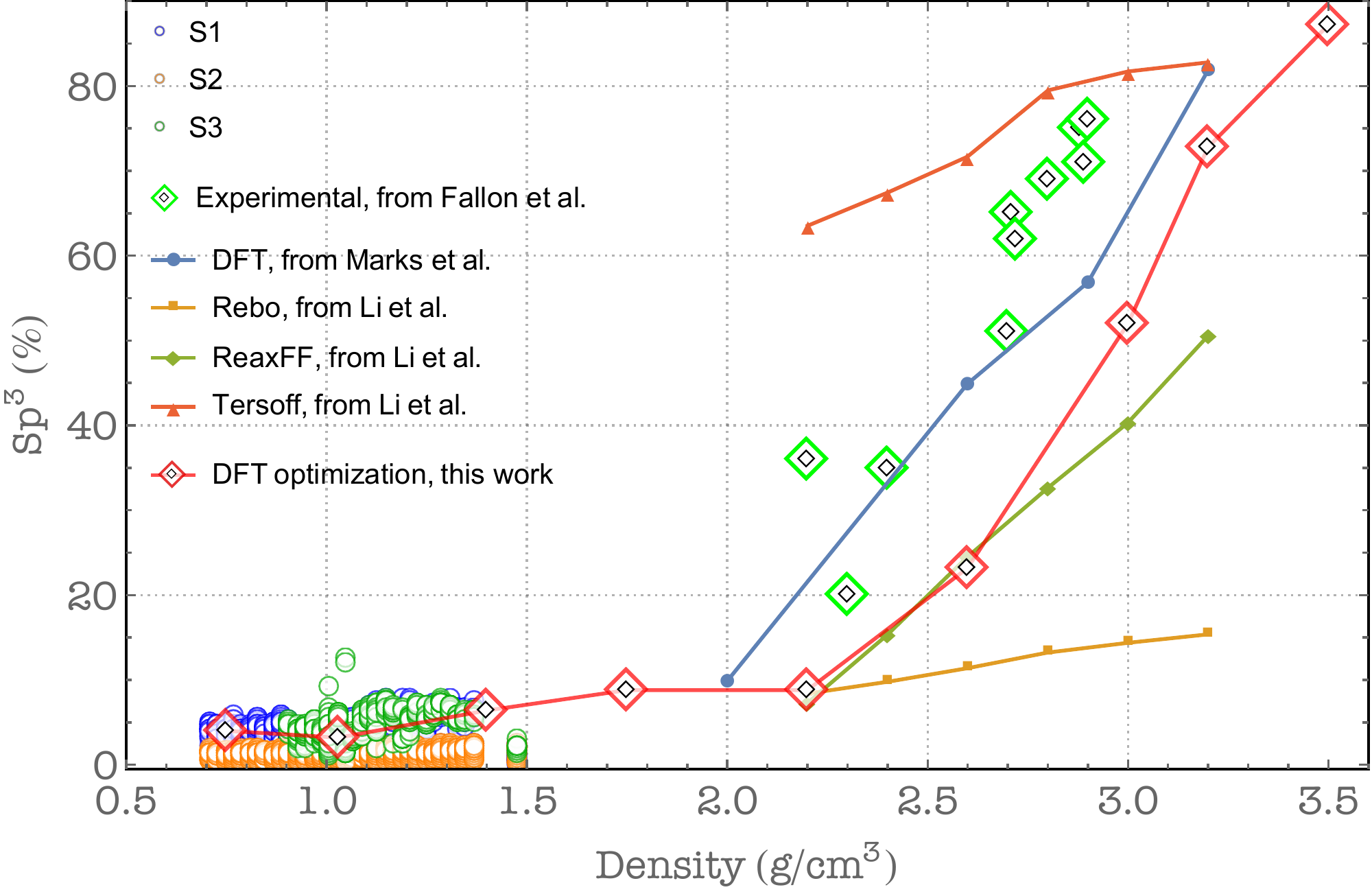}
		\end{center}
		\caption{\small\textit{Comparison between the results depicted in Li et. al with our classical and quantum results.}}
		\label{fig:14}
	\end{figure}

Based on Figure \ref{fig:10} related to the dependence of sp$^3$, we compared with the work of Li \textit{et. al} \cite{Li2013}. In their work, amorphous carbon was simulated  using the bond order of Tersoff potential and the reactive force fields 2dnREBO and ReaxFF. This comparison is shown in Figure \ref{fig:14}, where Li's  simulations were performed at densities higher than the triple point density and therefore, describe an amorphous phase with no porosity, far from our results. The authors concluded that Tersoff potential is suitable for densities close to diamond carbon, but not for low density structures, while 2nd REBO and ReaxFF are suitable potentials for low densities, those below 2.0~g/cm$^3$. In contrast, our results shows that Tersoff potential can be a suitable choice for the simulation for low density carbon; it can reproduce structures of nanoporous carbon at densities around 0.9~g/cm$^3$ as we showed in Figures \ref{fig:11},  \ref{fig:12} and \ref{fig:13} and the sp$^3$ concentration predicts a limit where all the simulations reside.

In addition, in figure \ref{fig:14} the stars represent our DFT relaxation calculation, described in section \ref{Sim}, obtained from cubic supercells, amorphized by our thermal procedure in LAMMPS and optimized, in the plane wave code QuantumESPRESSO. There, we have extended the density range of this simulations to observe the tendency in the high density range. It is remarkable that our DFT results are close to those of Marks \textit{et. al} \cite{Marks2002} made from Car-Parrinello molecular dynamics and applied to a quench-from-the-melt technique, in fact, around 2.0~g/cm$^3$ the two curves seems to coincide. All these results support the statement that Tersoff potential is a good choice to reproduce the atomic atomic structure of nanoporous carbon.

We can say that the many body features of the potential are sufficient conditions to generate, at first approximation, nanoporous amorphous carbon samples. The S1 set has two possible phases, one with higher energies, either graphite-like or spongy-like, and the other as spongy-like carbon. The S2 set has only spongy-like carbon while, S3 set is a phase with graphite-like nanoporous carbon.    

\section{Summary and Conclusions}
\label{Conc}

In this work we have produced an ensemble of 1700 nanoporous carbon structures using a heat-quench numerical procedure for densities between the critical and the triple point density, and the many-body interaction potential of Tersoff \cite{Tersoff1994}. The numerical procedure was applied to two precursors: unstable graphite supercells containing 6400 carbon atoms and unstable diamond supercells of 6480 carbon atoms, defining a sample space depicted in Figure \ref{fig:2}. To the resulted thermalized samples we measured the surface area, occupied volume, volume fraction, $sp^3$ content, radial distribution function, structures factor and simulated HRTEM images. Gathering all these information, we produced several metastable phase diagrams. In this work we presented the Volume fraction vs Surface area and the  $sp^3$ vs Pair energy diagrams. From the $sp^3$ vs Pair energy diagram we decided to apply a Gaussian-based clustering technique to automatically classify them. Three sets where found and labeled as S1, S2 and S3. In those sets we also found that: S1 presents a mix of spongy-like and a high energy phases with either graphite structure or spongy-like, both from the graphitic precursor. The S2 set contains entirely diamond precursors and forms a spongy-like carbon structure rich in $sp^2$ hybridization. Finally the S3 phase presents a low energy graphite-like structures that resembles nanoporous carbon as depicted in Figures \ref{fig:11}, \ref{fig:12} and \ref{fig:14}, which are comparisons between our simulated  and experimental results. 

From our results, we can conclude:

\begin{itemize}
	\item The NVT-CMD simulations applied to carbon Tersoff potential in unstable thermodynamic regions, placed between the critical and triple point, indeed lead to a bubble nucleation effect like the one observed in generic liquids. Practically any cooling rate applied to these states form nanoporous glasses.
	
	\item Despite its simplicity and the lack of a reactive nature, Tersoff potential can be used as a tool to generate nanoporous samples of carbon at low densities, at least densities ranging from 0.74 to 1.37~g/cm$^3$. The ensemble of structures produced by this CMD simulations can be characterized calculating properties as  surface area, volume fraction, potential energy, $sp^3$ content, among others, which are useful to explore structural phases, and lead to a reasonable good comparison with experimental measurements and other theoretical schemes.
	
	\item A diagram of sp$^3$ content vs potential energy, may be a good descriptor to find the structural phases. A clustering method may classify at least three different sets that were identified as layered graphitic-like, sponge-like and unstable structures.
	
	\item The application of the same heat-quench procedure to small systems to a further DFT relaxation, give results compatible with previous simulations and may be a good alternative to do intensive quantum studies on amorphous carbons, if more detailed characterizations are needed.
	
	\item Our present simulation scheme allow to produce very large sets of porous structures in reasonable computational times. Within these sets, data science based studies may be performed to propose new material design techniques for carbon-based devices. We successfully test the search of experimental structure factors in our base.
	
	\item The obtained structure data is also a source of nanoporous carbon structures where different complex phenomena can be studied at several length and time scales.
	
\end{itemize}

 \section*{Acknowledgement}
 
 We acknowledge the support given by DGAPA (Diercción General de Apoyo al Personal Académico) through project IN112414; 
 DGTIC (Direcci\'on General de C\'omputo y de Tecnolog\'ias de Informaci\'on y Comunicaci\'on) and the Supercomputing Department of Universidad Nacional Aut\'onoma de M\'exico for the computing resources under Project No. SC16-1-IR-29. On of us J.M. wants also to acknowledge the support given by C\'atedras-CONACYT (Consejo Nacional de Ciencia y Tecnolog\'ia) under Project No. 1191;

	\bibliographystyle{elsarticle-num}
	\bibliography{Article_LMMM.bbl}

\pagebreak

\end{document}